\documentclass[a4paper,12pt]{article}
\usepackage{latexsym,amssymb,amsfonts,amsmath,amsthm}
\usepackage[dvips]{graphicx}
\newtheorem{theorem}{Theorem}
\newtheorem{lemma}{Lemma}

\newtheorem{proposition}{Proposition}

\newtheorem{assumption}{Assumption}

\newcommand{\bs}[1]{\boldsymbol{#1}}
\newcommand{\im}{\bs{\rm i}}

\usepackage{color}
\newcommand{\segawa}{\textcolor{black}}
\newcommand{\HS}{\textcolor{black}}
\newcommand{\HSv}{\textcolor{black}}

\title{{\Large {\bf The spreading behavior of quantum walks \\ induced by 
drifted random walks on some magnifier graph  
}
}}
\author{ 
{\small 
Yusuke Higuchi,$^{1}$ 
\footnote{email address: higuchi@cas.showa-u.ac.jp
}\quad 
Etsuo Segawa,$^{2}$ 
\footnote{email address: e-segawa@m.tohoku.ac.jp 
}\quad
}\\ 
{\scriptsize $^{1}$ 
Mathematics Laboratories, College of Arts and Sciences, Showa University. 
}\\
{\scriptsize 
Fuji-Yoshida, Yamanashi 403-005, Japan. 
} \\
{\scriptsize $^3$ 
Graduate School of Information Sciences, Tohoku University, 
}\\
{\scriptsize 
Aoba, Sendai 980-8579, Japann
} \\
} 
\date{\empty }
\begin{document}
\maketitle

\par\noindent
\begin{small}
\par\noindent
{\bf Abstract}. 
In this paper, we consider the quantum walk on $\mathbb{Z}$ with attachment of one-length path periodically. 
This small modification to $\mathbb{Z}$ provides localization of the quantum walk. 
The eigenspace causing this localization is generated by finite length round trip paths. 
We find that the localization is due to the eigenvalues of an underlying random walk. 
Moreover we find that 
the transience of the underlying random walk provides a slow down of the pseudo velocity of the induced quantum walk 
and a different limit distribution from the Konno distribution. 


\end{small}

\setcounter{equation}{0}

\section{Introduction}
The Szegedy walk has been introduced~\cite{Sze} as an induced quantum walk (QW) by a random walk (RW). 
The spectrum of the Szegedy walk is decomposed into the inherited part by lifting up the spectrum of the underlying RW 
to the unit circle on $\mathbb{C}$, and its orthogonal complement subspaces named ``birth" part~\cite{Sze,Segawa}. 
Recently we show that the birth part of the Grover walk, which is a special case of the Szegedy walk, 
is related to a homological structure of the crystal lattice in \cite{HKSS}. 
This eigenspace gives a significant affect to the stochastic behavior of the walk: 
the positive valued first Betti number of the graph provides the non-empty birth eigenspace 
and ensures localization with an appropriate initial state. 
Here the appropriate initial state should have an overlap to the birth eigenspace. 
Characterizing the localization properties is one of the main topics in research on QW
; an answer is given in terms of geometrics of graph by \cite{HKSS}. 
There is a possibility that 
some previous works on localization of the Grover walks i.g., \cite{WKKK,SKJ1,SKJ2,IKS} and so on, 
can be inclusively shown by such a geometric explanation. 

From another point of view, we first show in this paper that the non-empty birth eigenspace is {\it not} a necessary condition for the localization. 
We find a crystal lattice which provides localization of the QW and has no cycles. 
The graph $G=(V(G),D(G))$ called the magnifier graph treated here is the maximal abelian covering graph of $M=(V(M),D(M))$. (see Fig.~\ref{fig:one}). 
Here $V(H)$ is the set of vertices and $D(H)$ is the set of symmetric arcs induced by edges of $H$, $(H\in \{G,M\})$. 
We denote by $\ell^2(D(G))$ the total state space of the QW with the standard inner product. 
The study of more general magnifier graph including our graph in this paper is studied in \cite{HS}. 
According to \cite{HS}, the spectrum of this graph has ``eigenvalues" and several number of absolutely continuous parts whose supports are mutually disjoint. 
One of our results implies that the eigenvalue of the underlying RW 
is another derivation of the localization in the induced QW. 
Instead of the cycles, the eigenspaces $\mathcal{L}_\pm$ defined by Eq.~(\ref{roundtrip})
are generated by finite length of round trip paths $\{p_j\}_{j\in \mathbb{Z}}$ in Eq.~(\ref{path}) (see also Fig.~\ref{fig:two}). 
While the Grover walk on one-dimensional lattice behaves the trivial one-way (strongly ballistic) walking, 
such a geometric modification provides a quite opposite property; localization to the original one:  
\begin{theorem}
Let $\mu_n: \mathbb{Z}\to [0,1]$ be the finding probability of the Grover walk at time $n$ in the $j$-th domain of $G$ defined by Eq.~(\ref{bunpu}). 
For any initial state $\Psi_0\in \ell^2(D(G))$, we have 
	\begin{align}\label{localization}
	\mu_n(j)\sim \sum_{e\in D(M)} | \langle \delta_{(j,e)},  (\Pi_{\mathcal{L}_+}+(-1)^n\Pi_{\mathcal{L}_-})\Psi_0 \rangle |^2. 
	\end{align}
Here for $\mathcal{H}'\subset \ell^2(D(G))$, $\Pi_{\mathcal{H}'}$ is the projection onto $\mathcal{H}'$. 
\end{theorem}

Let us consider the Szegedy walk on the one-dimensional lattice whose underlying random walker moves to right or left neighbors 
with probabilities $p$ and $1-p$, respectively. 
We see a duality between the underlying RW and the induced QW with respect to a spreading property: 
For a transient RW on the one-dimensional lattice, if the averaged distance at time $n$ from the starting position $|2p-1|n$ grows increasing, 
then the spreading strength of the corresponding Szegedy walk grows decreasing. 
See the following table. 
Here we evaluate the spreading strength of the QW as $\kappa\in [0,1]$ which is often called the pseudo velocity~\cite{MKK} as follows: 
\[ \kappa=\inf_{x\in\mathbb{R}} \{\lim_{n\to\infty}P(|X_n/n|<x)=1\}. \]
For the Szegedy walk on the one-dimensional lattice case, we have $\kappa=2\sqrt{p(1-p)}$. 
\begin{center}
\begin{tabular}{r|c|c}
  & RW & QW \\ \hline
$|2p-1|=0$ & recurrent & one-way ($\kappa=1$)\\
$0<|2p-1|<1$  & transient & $\kappa=2\sqrt{p(1-p)}$  \\
$|2p-1|=1$ & one-way & zigzag ($\kappa=0$) 
\end{tabular}
\end{center}
For $0<|2p-1|<1$, the limit behavior of the above Szegedy walk is well characterized by Konno~\cite{Konno,Konno2} as follows. 
Let $X_n^{(\mathbb{Z})}$ be a QW on the one-dimensional lattice at time $n$ with the mixed initial state. 
Then $X_n^{(\mathbb{Z})}/n$ weakly converges to named Konno's distribution \cite{Konno,Konno2} whose density is expressed by 
\begin{equation}\label{konno_density}
f_K(x;\kappa)=  \frac{\sqrt{1-\kappa^2}}{\pi (1-x^2)\sqrt{\kappa^2-x^2}}\bs{1}_{(-\kappa,\kappa)}(x), 
\end{equation}
where $\kappa=2\sqrt{p(1-p)}$. 
The above distribution described by Eq.~(\ref{konno_density}) frequently appears as a limit distribution of many kinds of QW models. 
See \cite{KLS} and its references therein. 
In this paper, we want to know how such properties of QW change on $G$, which is a ``little" modified one-dimensional lattice $\mathbb{Z}$. 

We construct a Szegedy walk so that both of the previous Grover walk and the Szegedy walk on one-dimensional lattice are included. 
This extension model is induced by the drifted RW characterized by three-independent parameters $(p,q,r)\in [0,1]^3$. 
See Fig.~\ref{fig:three} for the transition rule; 
the previous Grover walk corresponds to $(p,q,r)=(1/2,1/2,2/3)$ case, 
and the Szegedy walk on the one-dimensional lattice $X_n^{(\mathbb{Z})}$ corresponds to $(p,q,r)=(p,q,1)$ case. 
It is well known that the spectrum of simple RW is absolutely continuous and fills the whole of closed interval $[-1,1]$ if $G$ is just the one-dimensional lattice $\mathbb{Z}$. 
We see that the modification of the one-dimensional lattice causes a spectral gap of the RW $(-\lambda_0,\lambda_0)$ and the point spectrum at the origin. 
We also see that the transience of the RW causes the spectral gaps of the RW $(\lambda_1,1)$ and $(-1,\lambda_1)$, where $0\leq \lambda_0\leq \lambda_1\leq 1$. 
We show the following effects of these spectrum gaps on the spreading properties of the induced Szegedy walk: 
\begin{enumerate}
\item  a slow down of the pseudo velocity of the QW; 
\item  an additional different wave from the Konno density function in the weak limit theorem. 
\end{enumerate}
More precisely, we have the following theorem. 
\begin{theorem}
\noindent 
\begin{enumerate}
\item For any initial states, the pseudo velocity is expressed by 
\[ \kappa=\frac{1}{2}\sin(\theta_0-\theta_1). \]
Here $\theta_0=\arccos\lambda_0$ and $\theta_1=\arccos\lambda_1$. 
\item Let the walk start with the mixed initial state (see Assumption~1). Assume that $0<r<1$. Then we have 
\begin{multline}\label{weak_limit} 
	\frac{X_n}{n}\Rightarrow \frac{1}{3}\delta_0(x) \\
	+\frac{2}{3}\left\{ \gamma_-(x;\lambda_1,\lambda_0)+\gamma_+(x;\lambda_1,\lambda_0)\right\}
         \segawa{\times 2f_K\left(2x;\sqrt{1-\lambda_0^2}\right)} \mathbf{1}_{\{|x|<\kappa \}}(x), \;\;(n\to\infty). 
\end{multline}
Moreover the following four statements (a)-(d) are equivalent: 
	\begin{enumerate}
	\item The underlying RW is recurrent;
	\item $\lambda_1=1$;
	\item $\gamma_-(x;\lambda_1,\lambda_0)=0$, for all $|x|<\kappa$;
	\item $\gamma_+(x;\lambda_1,\lambda_0)=1$, for all $|x|<\kappa$. 
	\end{enumerate}
Here ``$Y_n \Rightarrow f(x)\;\;(n\to\infty)$" means that $\lim_{n\to\infty}\sum_{j<y}P(Y_n=j)=\int_{-\infty}^y f(x)dx$. 
See Eq.~(\ref{gamma}) for explicit expressions of $\gamma_-(x;\lambda_1,\lambda_0)$ and $\gamma_+(x;\lambda_1,\lambda_0)$. 
\end{enumerate}
\end{theorem}
For the Grover walk case, since the underlying RW is recurrent, then from the above theorem implies that 
the continuous part of the limit density function is described by $f_K(2x;\sqrt{1-\lambda_0^2})$. See Proposition.~\ref{principle}. 
We emphasize that in our model, such a situation that the limit density function is simply denoted by Konno's distribution is ``rare" in the following mean. 
In general, when the underlying RW is transient, that is, $p\neq q$, then 
the function $f_K(2x;\sqrt{1-\lambda_0^2})$ in Eq.~(\ref{weak_limit}) is cancelled by $\gamma_-(x,\lambda_1,\lambda_0)$ 
and the continuous part of the limit density of the induced QW is described by 
	\begin{equation*}
	\rho(x)=\frac{\sqrt{H_+(x)}}{\pi (1-4x^2)\sqrt{\phi(x)}}+\frac{\sqrt{H_-(x)}}{\pi (1-4x^2)\sqrt{\phi(x)}}. 
	\end{equation*}
Here $H_\pm(x)$ and $\phi(x)$ are expressed in terms of $\lambda_1$ and $\lambda_0$:
	\begin{multline*}
        H_{\pm}(x)=8\left\{-1+(1-\lambda_1^2)\lambda_0^2+(1-\lambda_0^2)\lambda_1^2 \right\}x^2 \\
        +2\left\{\lambda_1^2(1-\lambda_1^2)+\lambda_0^2(1-\lambda_0^2)\right\}\pm 2(\lambda_1^2+\lambda_0^2-1)\sqrt{\phi(x)}, 
        \end{multline*}
	\begin{equation*}
        \phi(x)=16x^4-8\left\{\lambda_1^2(1-\lambda_0^2)+\lambda_0^2(1-\lambda_1^2)\right\}x^2+(\lambda_1^2-\lambda_0^2)^2, 
        \end{equation*}
respectively. 

This paper is organized as follows. In Section 2, we present our settings of the graph and QWs treated here. 
In Section 3, we give the spectral mapping theorem from the underlying RW to the induced Szegedy walk. 
The proofs of Theorems 1 and 2 are provided in Sections 4 and 5, respectively. 
In the last half of Section 4 is the principal part of the Section 5. 
Differences between discussions Sections 4 and 5 make us understand the specificity of the Konno's distribution in our model. 
The basical idea of the proofs is based on the spectral mapping theorem of the twisted Szegedy walk 
from the underlying twisted RW~\cite{HKSS} on the Fourier space and the moment method of QWs~\cite{Grimmett} 
which is considered as useful to show the weak limit theorems of QWs. 
Finally, in Section 6, we give the summary. 
\section{Detailed settings}
\subsection{Definition of the magnifier graph}
Now we explain the graph treated here. Let $M=(V(M),E(M))$ be a finite graph defined by 
	\begin{align*}
	V(M)=\{R,S,T\},\;E(M)=\{RS,ST,ST\}, 
	\end{align*}
\HSv{where $E(M)$ is the set of unoriated edges.}  
Here $M$ has two edges whose end points are $S$ and $T$. 
We take the symmetric arcs $D(M)$ with respect to $E(M)$ as $D(M)=\{e_0,e_+,e_-\}\cup\{\overline{e}_0,\overline{e}_+,\overline{e}_-\}$, 
where $o(e_0)=o(e_+)=o(e_-)=S$, $t(e_0)=R$, $t(e_+)=t(e_-)=T$. 
Here for $e\in D(M)$, $o(e)$ and $t(e)$ are the origin and terminal vertices of $e$, respectively, and 
$\bar{e}$ is the inverse arc of $e$. 
The above $M$ coincides with a magnifier glass $M_{1,2}$ in \cite{HS}. 
Let $G$ be the maximal abelian covering graph $M^{ab}$. Here we call such a graph $G\cong M^{ab}$ of $M$; 
a {\it magnifier graph} under the identification $G$ with the quotient graph $M$. 
Then the graph $G$ is an infinite path to which the one-length paths are attached alternately. 
The graph $G$ can be regarded as $V(G)=\mathbb{Z}\times V(M)$ and $D(G)=\mathbb{Z}\times D(M)$. 
\begin{figure}[htbp]
\begin{center}
	\includegraphics[width=80mm]{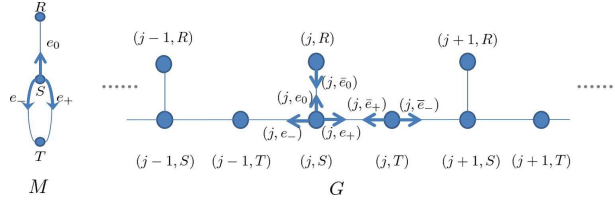}
\end{center}
\caption{The graph $M$ and the maximal abelian cover $G$: 
the vertex and arc fundamental domains are $\{(j,R),(j,S),(j,T)\}$ and $\{(j,e_0),(j,e_+),(j,e_-),(j,\bar{e}_0),(j,\bar{e}_+,\bar{e}_-)\}$, respectively. }
\label{fig:one}
\end{figure}
%
\subsection{Definition of the distribution for the induced QW}
In this paper, we consider the Grover walk and its extension model, the Szegedy walk, on $\ell^2(D(G))$; 
if the underlying RW is isotropic, the induced Szegedy walk is called the Grover walk. 
We consider a RW on $G$ given by the transition probability $p_G:D(G)\to (0,1]$ such that $\sum_{o(e)=u} p_G(u)=1$ for every vertex $u\in V(G)$. 
In other words, a particle at a vertex $o(e)$ moves to the vertex $t(e)$ along $e\in D(G)$ in a unit time with probability $p_G(e)$. 
If $p_G(e)=1/\mathrm{deg}(o(e))$, where $\mathrm{deg}(u)=\#\{e:o(e)=u\}$, then the random walk is called isotropic. 
At first, we consider the Grover walk whose underlying RW is isotropic. 
The induced Grover walk on $\ell^2(D(G))$ is defined as follows: 
the time evolution operator of the Grover walk $U: \ell^2(D(G))\to \ell^2(D(G))$ is denoted by for any $\psi\in \ell^2(D)$, 
	\begin{align*}
	(U\psi)(e)=\sum_{f:o(e)=t(f)}\left(\frac{2}{\mathrm{deg}(o(e))}-\delta_{e,\bar{f}}\right)\psi(f). 
	\end{align*}
Equivalently, for any $j\in \mathbb{Z}$, 
	\begin{align*}
	U\delta_{(j,e_0)} &= \delta_{(j,\bar{e}_0)}, \\
	U\delta_{(j,e_+)} &= \delta_{(j,\bar{e}_-)}, \;
	U\delta_{(j,e_-)} = \delta_{(j-1,\bar{e}_+)}, \\
        U\delta_{(j,\bar{e}_0)} &= -1/3 \cdot \delta_{(j,e_0)}+2/3 \cdot \delta_{(j,e_+)}+2/3 \cdot \delta_{(j,e_-)}, \\
        U\delta_{(j,\bar{e}_+)} &= 2/3 \cdot \delta_{(j,e_0)}-1/3 \cdot \delta_{(j,e_+)}+2/3 \cdot \delta_{(j,e_-)}, \\
        U\delta_{(j,\bar{e}_-)} &= 2/3 \cdot \delta_{(j+1,e_0)}+2/3 \cdot \delta_{(j+1,e_+)}-1/3 \cdot \delta_{(j+1,e_-)}.
	\end{align*}
Therefore at positions $(j,R), (j,T)\in \mathbb{Z}\times V(M)$, perfectly reflection and transmission happen, respectively. 
On the other hand, at position $(j,S)$, a non-trivial scattering happens. 
For given initial state $\Psi_0\in \ell^2(D(G))$ with $||\Psi_0||^2=1$ and $n\in \mathbb{N}$, because of the unitarity of $U$, 
we can define a probability distribution $\mu_n: \mathbb{Z}\to (0,1]$ as 
	\begin{equation}\label{bunpu}
	\mu_n(j)=\sum_{e\in D(M)} |\langle \delta_{(j,e)}, U^n\Psi_0\rangle|^2. 
	\end{equation}
This is interpreted as the probability that a quantum walker is found in the $j$-th domain 
$\{(j,e_0),(j,e_+),(j,e_-),(j,\bar{e}_0),(j,\bar{e}_+),(j,\bar{e}_-)\}$ at time $n$. 
In this paper, let a random variable $X_n$ follow $\mu_n$. 
Our interest is to clarify the asymptotic behavior of QW in terms of $X_n$. 
\subsection{Setting of the extension model}
\begin{figure}[htbp]
\begin{center}
	\includegraphics[width=80mm]{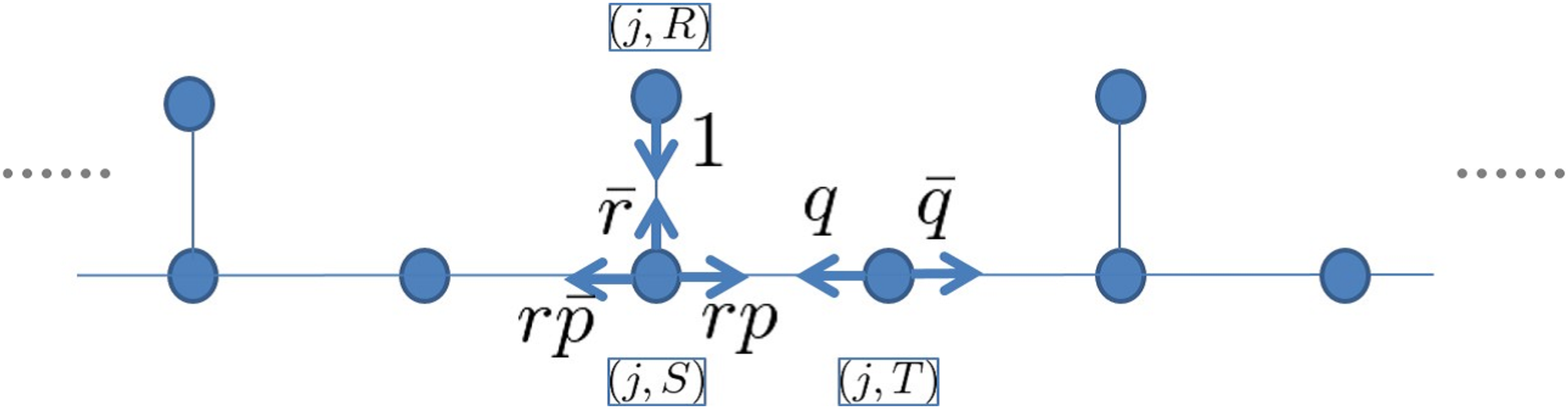}
\end{center}
\caption{The transition rule of the drifted underlying RW. }
\label{fig:three}
\end{figure}

Next, we extend the underlying RW as follows (see also Fig.~\ref{fig:three}): 
for any $f\in D(G)$ the transition probability $p_G: D(G)\to (0,1]$ is denoted by 
	\begin{equation}\label{extension}
	p_G(f)=
	\begin{cases}
	1 & \text{: $f\in \{(j,\bar{e}_0): j\in \mathbb{Z}\}$, }\\
        \bar{r} & \text{: $f\in \{(j,{e}_0): j\in \mathbb{Z}\}$, }\\
        rp & \text{: $f\in \{(j,{e}_+): j\in \mathbb{Z}\}$, }\\
        r\bar{p} & \text{: $f\in \{(j,{e}_-): j\in \mathbb{Z}\}$, }\\
        q & \text{: $f\in \{(j,\bar{e}_+): j\in \mathbb{Z}\}$, }\\
        \bar{q} & \text{: $f\in \{(j,\bar{e}_-): j\in \mathbb{Z}\}$, }
	\end{cases}
	\end{equation}
where $\bar{p}=1-p$, $\bar{q}=1-q$ and $\bar{r}=1-r$. 
The induced Szegedy walk on $\ell^2(D(G))$ is defined as follows: 
the time evolution operator of the Szegedy walk $U: \ell^2(D(G))\to \ell^2(D(G))$ is denoted by for any $\psi\in \ell^2(D)$, 
	\begin{align*}
	(U\psi)(e)=\sum_{f:o(e)=t(f)}\left(2\sqrt{p(e)p(\bar{f})}-\delta_{e,\bar{f}}\right)\psi(f). 
	\end{align*}
In the above Szegedy walk, to the arc $e$, the transmission rate from the arc $f(\neq \bar{e})$ with $o(e)=t(f)$ is $2\sqrt{p(e)p(\bar{f})}$ while the 
reflection rate from the inverse arc $\bar{e}$ is $2|p(e)|^2-1$.

\section{Spectral property}
In the following discussion, we assume $0<p,q,r<1$. 
Let $P$ be the transition operator of the RW on $G$ equipped with Eq.~(\ref{extension}). 
Since $G$ has no non-trivial cycles, such a RW is always reversible, that is, there exists a positive valued function $m: V(G)\to \mathbb{R}^+$ such that 
	\[ p_G(e)m(o(e))=p_G(\bar{e})m(t(e))\;\;(e\in D(G)). \]
Actually we can select $m(\cdot)$ as 
	\begin{align*} 
          m((j,S)) &= \left(\frac{p\bar{q}}{\bar{p}q}\right)^j, \\
          m((j,T)) &= \frac{rp}{q}m((j,S)),\;
          m((j,R))=\bar{r}m((j,S)). 
        \end{align*}
It is well known that the reversible transition operator $P$ on $\ell^2(V,m)$ and its symmetrized operator $J$ on $\ell^2(V)$ are unitarily equivalent; thus we focus on the spectrum of $J$. 
Let $P$ be the transition operator of the RW in the setting of Eq.~(\ref{extension}) and $J$ its symmetrized operator, that is, 
$(J)_{u,v}=\sqrt{(P)_{u,v}(P)_{v,u}}=\sqrt{p_G(e)p_G(\bar{e})}$, $(o(e)=v,\;t(e)=u)$. 
Denote $f_n\in \ell^2(V)$ as the $n$-th iteration of $J$, $f_n=Jf_{n-1}$. 
\HSv{We set a $1$-form $\theta: D(M)\to \mathbb{R}$ such that 
	\begin{equation}\label{oneform}
        \theta(e)=\begin{cases} -k & \text{: $e=e_-$,} \\ k & \text{: $e=\bar{e}_-$,}\\ 0 & \text{: $otherwise$.} \end{cases}
        \end{equation}
See Fig.~\ref{fig:one} for the fundamental domain. 
We take $p_M: D(M)\to (0,1]$ by $p_M(e)=p_G((j,e))$. }
Taking the Fourier transform $\mathcal{F}^*:\ell^2(\mathbb{Z}\times V(M))\to L^2((0,2\pi]\times V(M))$, that is, 
\[(\mathcal{F}^*f)(k;u)\equiv \hat{f}_n(k;u)=\sum_{j\in \mathbb{Z}}f_n(j,u)e^{\im kj}\;\; (k\in[0,2\pi)), \] 
we have 
	\begin{equation*}
	^T[\hat{f}_n(k;R),\hat{f}_n(k;S),\hat{f}_n(k;T)] =J_k\;^T[\hat{f}_{n-1}(k;R),\hat{f}_{n-1}(k;S),\hat{f}_{n-1}(k;T)], 
	\end{equation*}
where $\im=\sqrt{-1}$. 
\HSv{Here the transition matrix of the twisted RW $J_k: \ell^2(V(M))\to \ell^2(V(M))$ is denoted by 
	\begin{multline*} 
        (J_kg)(u)=\sum_{f:t(f)=u}e^{\im \theta(f)} \sqrt{p_M(f)p_M(\bar{f})} g(o(f))\\ (g\in \ell^2(V(M))). 
        \end{multline*}
More precisely, $J_{k}$ can be represented by}
	\HS{\begin{equation*}
        J_k=
        \begin{bmatrix}
        0 & \sqrt{\bar{r}} & 0 \\
        \sqrt{\bar{r}} & 0 & \sqrt{r\bar{p}\bar{q}}e^{\im k}+\sqrt{rpq} \\
        0 & \sqrt{r\bar{p}\bar{q}}e^{-\im k}+\sqrt{rpq} & 0
        \end{bmatrix}.
        \end{equation*}}
We have the spectrum of $J_k$ as 
	\[ \mathrm{spec}(J_k)=\left\{0, \pm \sqrt{a+b\cos k}\right\}, \]
where we put $a=a(p,q,r)=1-r(p\bar{q}+q\bar{p})$ and $b=b(p,q,r)=2r\sqrt{\bar{p}\bar{q}pq}$. 
Thus we can compute that 
	\begin{equation}\label{specP1}
        \mathrm{spec}(J)= 
        [-\lambda_1,-\lambda_0]\cup \{0\} \cup [\lambda_0,\lambda_1], 
        \end{equation}
where 
	\begin{align}
	\lambda_0^2 & = 1-r(\sqrt{p\bar{q}}+\sqrt{\bar{p}q})^2, \label{pero}\\ 
	\lambda_1^2 & =1-r(\sqrt{p\bar{q}}-\sqrt{\bar{p}q})^2. \label{pipi}
	\end{align}
Note that $0<1-r\leq \lambda_0^2<\lambda_1\leq 1$. Here the first equality holds if and only if $p+q=1$, and the second equality holds if and only if $p=q$. 
From a simple observation, we obtain the following lemma. 
\begin{lemma} The following statements are equivalent:
\begin{enumerate}
\item Random walk on $G$ is recurrent;
\item $\lambda_1=1$; 
\item $p=q$. 
\end{enumerate}
\end{lemma}
In parallel, we define the time evolution of a twisted Szegedy walk $\widehat{U}_k=S_kC$ on $\ell^2(D(M))$. 
	\begin{align*}
	S_k\psi (e) &= e^{-\im\theta(e)}\psi(\bar{e}), \\
	C\psi (e) &= \sum_{f:o(f)=o(e)} \left( 2\sqrt{p_G(e)p_G(e)}-\delta_{e,f}\right)\psi(\bar{f}). 
	\end{align*}
\HSv{Here the $1$-form $\theta$ is same as in Eq.~(\ref{oneform}). }
By the spectral mapping theorem \HSv{between RW and QW~\cite{HKSS}}, spectra of the twisted Szegedy walk on $\ell^2(D(M))$ is described by 
	\begin{equation}\label{specmap} 
        \mathrm{spec}(\widehat{U}_k)=\varphi_{QW}^{-1}\left(\left\{0, \pm \sqrt{a+b\cos k}\right\}\right), 
        \end{equation}
where $\varphi_{QW}(x)=(x+x^{-1})/2$. 
We define the spatial Fourier transform $\mathcal{F}^*: \ell^2(\mathbb{Z}\times D(M))\to L^2([0,2\pi)\times D(M))$ such that 
	\[ (\mathcal{F}^*\psi)(k,f)=\sum_{x\in \mathbb{Z}}\psi(x,f)e^{\im kx}, \]
and the inverse Fourier transform $\mathcal{F}: L^2([0,2\pi)\times D(M))\to \ell^2(\mathbb{Z}\times D(M))$, such that 
	\[ (\mathcal{F}g)(x,f)=\int_0^{2\pi}g(k,f)e^{-\im kx}\frac{dk}{2\pi}. \]
We should remark that for any $n\in\mathbb{N}$, 
	\[ \mathcal{F}^*(U^n \Psi_0)=\widehat{U}_k^n\mathcal{F}^*(\Psi_0).  \]
Taking $\widehat{\Psi}_n(k)=\widehat{U}_k^n\mathcal{F}^*(\Psi_0)$, it holds that
	\begin{equation}\label{itsumono} 
	E[e^{\im\xi X_n}]=\int_{0}^{2\pi} \langle \widehat{\Psi}_n(k), \widehat{\Psi}_n(k+\xi) \rangle\frac{dk}{2\pi}. 
	\end{equation}
The above equation is an extended expression of the moment method of QWs \cite{MKK, Grimmett}. 
Equation (\ref{specmap}) implies that 
	\[\mathrm{spec}(\widehat{U}_k)=\{\pm \im,\;\pm e^{\im\arccos\sqrt{a+b\cos k}}, \pm e^{-\im\arccos\sqrt{a+b\cos k}}\}.  \]  
\section{Grover walk case}
\subsection{Proof of Theorem~1}

\begin{figure}[htbp]
\begin{center}
	\includegraphics[width=60mm]{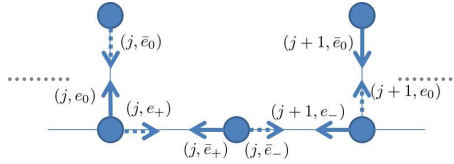}
\end{center}
\caption{A graphical representation of the round trip path $p_j=(q_j,\overline{q_j})$. The real and dotted lines depict $q_j$ and $\overline{q_j}$, respectively. }
\label{fig:two}
\end{figure}

We introduce a round trip path $p_j=(q_j,\overline{q_j})$ in $G$, where 
	\begin{equation}\label{path} 
        q_j =(f_1^{(j)},f_2^{(j)},f_3^{(j)},f_4^{(j)}),\; \overline{q_j} =(\bar{f}_4^{(j)},\bar{f}_3^{(j)},\bar{f}_2^{(j)},\bar{f}_1^{(j)}).
        \end{equation} 
Here $f_1^{(j)}=(j,\bar{e}_0)$, $f_2^{(j)}=(j,e_+)$, $f_3^{(j)}=(j,\bar{e}_-)$ and  $f_4^{(j)}=(j+1,e_0)$. 
Refer to Fig.~\ref{fig:two}. 
Moreover we define $w: \{p_j: j\in \mathbb{Z}\}\to \ell^2(D(G))$, which plays an important role to describe localization, by 
	\[ w(p_j)=\sum_{m=1}^{4} r_m (-\im)^{m-1} \left(\delta_{f_m^{(j)}}+\im \delta_{\bar{f}_m^{(j)}} \right), \] 
where $r_1=\sqrt{rpq}$, $r_2=\sqrt{q\bar{r}}$, $r_3=\sqrt{\bar{r}\bar{q}}$, $r_4=\sqrt{\bar{q}\bar{p}r}$. 
Thus it is obvious that $w(p_j)$ has a finite support, and $w(p_j)$ and $w(p_k)$ are linearly independent $(j\neq k)$. 
Denote $\mathcal{L}_\pm\subset \ell^2(D(G))$ by 
	\begin{equation}\label{roundtrip} 
        \mathcal{L}_+=\mathrm{span}\left\{w(p_j):j\in \mathbb{Z}\right\},\;
	\mathcal{L}_-=\mathrm{span}\left\{\overline{w(p_j)}:j\in \mathbb{Z}\right\}. 
        \end{equation}

The result on the localization of Theorem~1 in terms of the Grover walk, whose parameter is $(p,q,r)=(1/2,1/2,2/3)$, can be extended to the Szegedy walk as follows. 

\noindent \\
\noindent{\bf Theorem~1$'$} 
{\it For the Szegedy walk $(p,q,r)\in(0,1)^3$ with every initial state $\Psi_0\in \ell^2(D(G))$, we have 
	\begin{align}\label{localization}
	\mu_n(j)\sim \sum_{e\in D(M)} | \langle \delta_{(j,e)},  (\Pi_{\mathcal{L}_+}+(-1)^n\Pi_{\mathcal{L}_-})\Psi_0 \rangle |^2. 
	\end{align}
Here for $\mathcal{H}'\subset \ell^2(D(G))$, $\Pi_{\mathcal{H}'}$ is the projection onto $\mathcal{H}'$. }
\noindent \\

In the rest of this subsection, we give the proof of Theorem~1$'$. 
We introduce a boundary operator $d_A: \ell^2(D(M))\to \ell^2(V(M))$ such that 
	\[(d_A\psi)(u)= \sum_{e:o(e)=u}\sqrt{p(e)}\psi(e). \]
Also we introduce its adjoint operator $d_A^*: \ell^2(V(M))\to \ell^2(D(M))$ such that $(d_A^*f)(e)=\sqrt{p(e)} \cdot f(o(e))$. 
For fixed $k\in[0,2\pi)$, according to \cite{HKSS}, for every eigenvalue $e^{\im\nu}\in \mathrm{spec}(\widehat{U}_k)\setminus \{\pm 1\}$, 
its eigenfunction $\psi\in \ell^2(D(M))$ can be expressed by using the eigenfunction $f\in \ell^2(V(M))$ of eigenvalue $\cos \nu\in \mathrm{spec}(J_k)\setminus\{\pm 1\}$: 
	\begin{equation}\label{eigenvectors} 
        \psi=(I-e^{-\im\nu}S_k)d_A^*f.  
        \end{equation}
By the inverse Fourier transform, taking $\Psi_n: \mathbb{Z}\to \ell^2(D(M))$ such that $(\Psi_n(j))(e)=\langle \delta_{(j,e)}, U^n\Psi_0\rangle$ 
	\begin{align}\label{inverseF}
        \Psi_n(j) = \int_{0}^{2\pi} \widehat{\Psi}_n(k)e^{-\im kj}\frac{dk}{2\pi}. 
	\end{align}
Taking $\nu(k)=\arccos \sqrt{a+b\cos k}\in \arccos(\mathrm{spec}(J_k))$, 
we can decompose $\widehat{U}_k$ into 
	\begin{equation}\label{decom}
        \widehat{U}_k = \im \Pi_{v_+} -\im \Pi_{v_-} 
        + e^{\im \nu(k)} (\Pi_{u_{++}}-\Pi_{u_{+-}})
        + e^{-\im \nu(k)} (\Pi_{u_{-+}}-\Pi_{u_{--}}). 
        \end{equation}
Here $v_\pm$, $u_{\epsilon\tau} \in \ell^2(D(M))$ $(\epsilon,\tau\in\{\pm\})$ are the normalized eigenfunctions of the eigenvalues $\pm \im$, $\tau e^{\epsilon\im\nu(k)}$, respectively, and 
$\Pi_w$ is the projection onto $w\in \ell^2(D(M))$. 
Thus, the Riemann-Lebesgue Lemma implies that the contribution of the eigenvalues $\tau e^{\epsilon \im\nu(k)}$ is vanishes for $n\to\infty$: 
        \begin{align}\label{RL}
        \Psi_n(j) \sim \im^n \int_0^{2\pi} \left\{\Pi_{v_+} + (-1)^n \Pi_{v_-} \right\}e^{-\im kj}  \widehat{\Psi}_0  \frac{dk}{2\pi}. 
	\end{align}
An explicit expression for the eigenfunction of eigenvalue $0\in \mathrm{spec}(J_k)$ is 
	\[ f_0={}^T\begin{bmatrix}\sqrt{\bar{p}\bar{q}r}e^{\im k}+\sqrt{rpq} & 0 & -\sqrt{\bar{r}}\end{bmatrix}. \]
Using Eq.~(\ref{eigenvectors}), we have the corresponding eigenfunctions $v_\pm\in \ell^2(D(M))$ 
	\[ v_\pm= c^{-1}(I\pm \im S_k) d_A^* f_0. \]
Here $c\in \mathbb{C}$ is a normalized constant. 
Therefore the eigenvector on $\ell^2(\mathbb{Z},D(M))$ lifted from $L^2([0,2\pi), D(M))$ is expressed by
	\begin{equation}\label{kurukuru} 
        \mathcal{F}(ce^{\im jk} v_+)=w(p_j),\;\mathcal{F}(ce^{\im jk} v_-)=\overline{w(p_j)}, 
        \end{equation}
for $j\in \mathbb{Z}$. 
Combining Eq~(\ref{RL}) with Eq.~(\ref{kurukuru}) implies
	\begin{equation*}
        (\Psi_n(j))(e) \sim \im^n \langle \delta_{(j,e)}, (\Pi_{\mathcal{L}+}+(-1)^n\Pi_{\mathcal{L}-})\Psi_0\rangle, 
        \end{equation*}
which leads to our desired conclusion. $\square$

\subsection{Principal part of the proof of the weak limit theorem}
For simplicity, let us consider the following initial state.  
\begin{assumption}
We take the initial state $\Psi_0\in \ell^2(D(G))$ provided uniformly from $\{ \delta_{(0,e)}: e\in D(M) \}$; that is,
	\[ P(\Psi_0=\delta_{(o,e)})=1/6,\;\; (e\in D(M)). \]
We call the initial state ``mixed initial state''. 
\end{assumption}
\begin{lemma}
If the initial state is the mixed state, then in the Szegedy walk case, we have 
	\begin{equation}\label{kyokugen}
        \lim_{n\to \infty}E[e^{\im\xi X_n/n}]= \frac{1}{3} 
        	+\frac{1}{3}\int_{0}^{2\pi} \exp[\im\xi \nu'(k)]\frac{dk}{2\pi}+\frac{1}{3}\int_{0}^{2\pi}\exp[-\im\xi \nu'(k)] \frac{dk}{2\pi}. 
        \end{equation}
\end{lemma}
{\it Proof. }
Replacing $\xi$ to $\xi/n$ in Eq.~(\ref{itsumono}), 
	\begin{multline*}
        \langle \widehat{\Psi}_n(k), \widehat{\Psi}_n(k+\xi/n) \rangle = 
        \mathrm{Tr}[(\Pi_{v_+}+\Pi_{v_-})\widehat{\Psi}_0\widehat{\Psi}_0^\dagger ] \\
                + e^{\im\xi \nu'(k)}\mathrm{Tr}[(\Pi_{u_{++}}+\Pi_{u_{+-}})\widehat{\Psi}_0\widehat{\Psi}_0^\dagger] \\
                +e^{-\im\xi \nu'(k)}\mathrm{Tr}[(\Pi_{u_{--}}+\Pi_{u_{-+}})\widehat{\Psi}_0\widehat{\Psi}_0^\dagger ]
                +O(n^{-1}).
        \end{multline*}
Here $\nu'(k)=\partial \nu(k)/\partial k$. 
Notice that $E[\widehat{\Psi}_0\widehat{\Psi}_0^\dagger]= 1/6\cdot I$, which implies that 
	\begin{align*}
        \mathrm{Tr}[(\Pi_{v_+}+\Pi_{v_-})\widehat{\Psi}_0\widehat{\Psi}_0^\dagger ]
        	&= \mathrm{Tr}[(\Pi_{u_{++}}+\Pi_{u_{+-}})\widehat{\Psi}_0\widehat{\Psi}_0^\dagger] \\
         	&= \mathrm{Tr}[(\Pi_{u_{--}}+\Pi_{u_{-+}})\widehat{\Psi}_0\widehat{\Psi}_0^\dagger] \\
                &=1/3. 
	\end{align*}
Thus taking $n\to \infty$, we can represent Eq.~(\ref{itsumono}) by Eq.~(\ref{kyokugen}). $\square$

To see the outline of the proof of the weak limit theorem, we treat the Grover walk case, that is, $r=2/3$, $p=q=1/2$ in the rest of this subsection. 
More general case can be seen in the next section. 
Taking the time average of Eq.~(\ref{localization}); $ \overline{\mu}_\infty(j)\equiv  \lim_{T\to\infty}\frac{1}{T} \sum_{n=0}^{T-1}\mu_n(j)$, we have 
	\[ \sum_{j\in \mathbb{Z}}\overline{\mu}_\infty(j)=\frac{1}{3}. \]
So $\overline{\mu}_\infty$ is no longer probability distribution. 
We can recover the missing value $1-1/3=2/3$ by taking scaling linearly as follows: 
\begin{proposition}\label{principle}
Assume that the initial state is provided by Assumption~1. In the Grover walk case, we have 
 	\begin{equation}
	\frac{X_n}{n}\Rightarrow \frac{1}{3}\times \delta_{0}(x)+\frac{2}{3}\times 2f_K(2x;\sqrt{2/3}).
         \\(n\to\infty)
	\end{equation}
\end{proposition}
{\it Proof. }
To obtain the limit density function of $X_n/n$, we replace $\nu'(k)$ to $x$ in the integrands of second and last terms in RHS of Eq.~(\ref{kyokugen}). 
To do so, we compute the Hessian of $\nu(k)$ by taking $x=\nu'(k)$. Recall that $\cos \nu(k)=\sqrt{(2+\cos k)/3}$ which implies 
	\begin{equation}\label{three_state_atode}
	\cos [2\nu(k)]=\frac{1+2\cos k}{3}.
	\end{equation}
Taking differential with respect to $k$ to both sides, we have 
	\begin{equation}\label{sol1}
        x=\frac{\sin k}{3\sin [2\nu(k)]}. 
        \end{equation}
Obviously, the variable $x$ is bounded as follows: 
	\begin{equation}\label{sol2} 
        x^2=\frac{1+\cos k}{4 (2+\cos k)}\leq \frac{1}{6}.  
        \end{equation}
This is the support of the limit density function. Equation (\ref{sol2}) implies the inverse function of Eq.~(\ref{sol1}):
	\begin{align}\label{inverseSol}
        \cos k=\frac{8x^2-1}{1-4x^2}.
	\end{align} 
Again taking differential with respect to $k$ to both sides of Eq.~(\ref{sol1}), we obtain
	\begin{align}\label{hessian}
        \frac{\partial^2 \nu(k)}{\partial k^2}=\frac{\cos k-6x^2 \cos [2\nu(k)]}{3\sin [2\nu(k)]}.
        \end{align}
Combining Eq.~(\ref{hessian}) with Eq.~(\ref{inverseSol}), we arrive at the Hessian of $\nu(k)$: 
	\begin{align}\label{HessianX}
        \bigg| \frac{\partial^2 \nu(k)}{\partial k^2} \bigg|^{-1} = \frac{\sqrt{8}}{(1-4x^2)\sqrt{1-6x^2}}.
        \end{align}
Therefore inserting Eq.~(\ref{HessianX}) into Eq.~(\ref{kyokugen}) provides 
	\begin{equation}\label{conc}
        \lim_{n\to \infty}E[e^{\im\xi X_n/n}]= \frac{1}{3}+\frac{2}{3}\int_{-\infty}^{\infty} e^{\im\xi x} 2f_K(2x;\sqrt{2/3})dx. 
        \end{equation}
The proof is completed. $\square$
\section{Szegedy walk case}
\subsection{Proof of part 1 of Theorem 2}
\HS{In the Grover walk case, we can immediately obtain the inequality Eq.~(\ref{sol2}) from Eq.~(\ref{sol1}). On the contrary, in the Szegedy walk case, we have to do somewhat complicated discussion 
in order to obtain the corresponding inequality. }

Let us recall Eq.~(\ref{kyokugen}): 
	\begin{equation*} 
        \lim_{n\to \infty}E[e^{\im\xi X_n/n}]= 
	\frac{1}{3} 
        +\frac{1}{3}\int_{0}^{2\pi} \exp[\im\xi \nu'(k)]\frac{dk}{2\pi} +\frac{1}{3}\int_{0}^{2\pi}\exp[-\im\xi \nu'(k)] \frac{dk}{2\pi}.  
        \end{equation*}
Here remark that $\cos\nu(k)=\sqrt{a+b\cos k}$. 
Taking differential with respect to $k$, we have 
	\begin{equation}\label{ikkaibibun} 
        x=\frac{\partial \nu(k)}{\partial k}= \frac{b\sin k}{\sin 2\nu (k)}. 
        \end{equation}
From now on we compute the maximal value of $x^2$. We can rewrite $x^2$ by 
	\begin{equation}\label{cos} 
        x^2=\frac{1}{4}\;\frac{(1-2\alpha+\cos2\nu(k))(1-2\beta+\cos2\nu(k))}{\cos^22\nu(k)-1}, 
        \end{equation}
where $\alpha\equiv a+b=\lambda_1^2$, $\beta\equiv a-b=\lambda_0^2$. 
So replacing $\cos 2\nu(k)$ with $t$, 
we can regard RHS as the function of $f(t)$ with $2\beta-1\leq t\leq 2\alpha-1$: 
	\begin{equation}\label{higuchi}
        f(t)=\frac{1}{4}\;\frac{(1-2\alpha+t)(1-2\beta+t)}{t^2-1}\geq 0.
        \end{equation}
The derivative is computed by
	\begin{equation*}
        f'(t)=\frac{(\alpha+\beta-1)t^2-2\{1-(\alpha+\beta)+2\alpha\beta\}t+(\alpha+\beta-1)}{2(t^2-1)^2}. 
        \end{equation*}
The zero, $t_*$, of the derivative between $2\beta-1$ and $2\alpha-1$ comes from its numerator. 
When $\alpha+\beta-1=0$, then $t_*=0$, which implies $0\leq f(t)\leq f(0)=(1-2\alpha)^2/4$. 
\HS{Thus we have $|x|\leq (1/2) \cdot \sin(\theta_0-\theta_1)$ since $\alpha+\beta=\lambda_0^2+\lambda_1^2$. }

Now we assume $\alpha+\beta-1\neq 0$. It holds that $f'(t)=0$ if and only if 
	\begin{equation}\label{nijieq}
        t^2+2\left(1-\frac{2\alpha\beta}{\alpha+\beta-1}\right)t+1=0.
	\end{equation}
The solutions for Eq.~(\ref{nijieq}), say $t_1$ and $t_2$, can be assumed that $|t_1|\leq 1 \leq |t_2|$ since $t_1t_2=1$. 
We can easily check that $t_1=t_*$ and $t_2=1/t_*$ as 
	\begin{align*} 
        t_* &= \frac{1-(\sqrt{\alpha(1-\beta)}+\sqrt{\beta(1-\alpha)})^2}{\alpha+\beta-1},\\
        1/t_* &= \frac{1-(\sqrt{\alpha(1-\beta)}-\sqrt{\beta(1-\alpha)})^2}{\alpha+\beta-1} 
        \end{align*}
and that $-1\leq 2\beta-1 \leq t_*\leq 2\alpha-1\leq 1$. 
By Eq.~(\ref{nijieq}), we have 
	\[ (1-2\alpha)(1-2\beta)=(\alpha+\beta-1)(t_*+1/t_*)-1, \]
which implies that the numerator of Eq.~(\ref{higuchi}) for $t=t_*$ is expressed by 
	\[ (t_*^2-1)\left(1-\frac{\alpha+\beta-1}{t_*} \right). \]
Combining the fact $f(2\alpha-1)=f(2\beta-1)=0$ with the above expression of the numerator of $f(t_*)$,  
we have 
	\[ 0\leq f(t)\leq f(t_*)=\left(\frac{\sin (\theta_0-\theta_1)}{2}\right)^2. \] 
The proof is completed. $\square$
\subsection{Proof of part 2 of Theorem 2}
While the limit distribution Eq.~(\ref{conc}) follows from Eq.~(\ref{hessian}) immediately in the Grover walk case, we have to treat and investigate more complicated functions to get 
the corresponding distribution in the Szegedy walk case. Such a complicated computation brings us an interesting limit distribution consisting of two waves. 

Now let us compute an explicit expression for the density function. 
We take differential with respect to $k$ to both sides of Eq.~(\ref{ikkaibibun}) again, we have 
	\begin{equation}\label{nikaibibun}
        \frac{\partial^2 \nu(k)}{\partial k^2}=\frac{1-(\alpha+\beta)}{2\sin 2\nu(k)}+\frac{1}{2}(1-4x^2)\frac{\cos 2\nu(k)}{\sin2\nu(k)}. 
	\end{equation}
Here we put $x=\partial \nu(k)/\partial k\in (-\sin(\theta_0-\theta_1)/2, \sin(\theta_0-\theta_1)/2)$. 
The target is to obtain the closed form of the RHS of Eq.~(\ref{nikaibibun}) with respect to $x$. 
We should remark that when $\alpha=1$, then the term $(1-t)$ appears in the numerator of RHS of Eq.~(\ref{cos}). 
Eq.~(\ref{cos}) is reduced to 
	\[ x^2=-\frac{1}{4}\frac{1-2\beta+t}{t+1}, \]
where $t=\cos2\nu(k)$. 
Thus $\cos^2 2\nu(k)$ can be uniquely determined if $\alpha=1$. 
On the other hand, if $\alpha\neq 1$, the inverse function $f^{-1}(x)$ becomes a multiple-valued function with respect to Eqs.~(\ref{cos}) or (\ref{higuchi}). 
This fact implies the two waves in the weak limit theorem for {\it transient} underlying RW. 
The following is the expression for $f^{-1}(x^2)$. 
	\begin{equation}\label{cos2}
        f^{-1}(x^2)=\cos 2\nu(k)=\frac{1-(\alpha+\beta)\pm \sqrt{\phi(x)}}{4x^2-1}\equiv h_{\pm}(x), 
        \end{equation}
where 
	\begin{equation}\label{phi}
        \phi(x)=16x^4-8\left\{\lambda_1^2(1-\lambda_0^2)+\lambda_0^2(1-\lambda_1^2)\right\}x^2+(\lambda_1^2-\lambda_0^2)^2. 
        \end{equation}
Recall $\alpha=\lambda_1^2$ and $\beta=\lambda_0^2$. 
It holds that $\phi(\pm \kappa)=0$ and $\phi(x)\geq 0$ for $|x|\leq \kappa$. 
The integrands of second term of RHS in Eq.~(\ref{kyokugen}) is changed to 
	\begin{equation}\label{jacobi} 
        e^{\im\xi \nu'(k)}dk 
        =e^{\im\xi x}\bigg(\frac{1}{|\nu''(k)|}\bigg|_{\cos 2\nu(k)=h_+(x)}
        +\frac{1}{|\nu''(k)|}\bigg|_{\cos 2\nu(k)=h_-(x)}\bigg)dx. 
        \end{equation}
Directly inserting Eq.~(\ref{cos2}) into Eq.~(\ref{nikaibibun}), we have 
	\begin{equation}\label{hessian2} 
        \frac{1}{|\nu''(k)|}\bigg|_{\cos 2\nu(k)=h_\pm (x)}=\frac{2\sqrt{H_\pm (x)}}{(1-4x^2)\sqrt{\phi(x)}}, 
        \end{equation}
where 
	\begin{multline}\label{H}
        H_{\pm}(x)=8\left\{-1+(1-\lambda_1^2)\lambda_0^2+(1-\lambda_0^2)\lambda_1^2 \right\}x^2 \\
        +2\left\{\lambda_1^2(1-\lambda_1^2)+\lambda_0^2(1-\lambda_0^2)\right\}\pm 2(\lambda_1^2+\lambda_0^2-1)\sqrt{\phi(x)}. 
        \end{multline}
This is the Hessian for changing variable by $x=\nu'(k)$ which is generally different from $f_K$ defined in Eq.~(\ref{konno_density}). 
Substituting Eq.~(\ref{hessian2}) into Eq.~(\ref{jacobi}), we get the explicit expression for the density function. 
To extract an effect on the transient behavior of the underlying RW more clearly, 
we can rewrite Eq.~(\ref{hessian2}) after the substitution of Eq.~(\ref{jacobi}) as 
	\begin{equation}\label{hessian4} 
        \left\{\gamma_+(x;\lambda_1,\lambda_0)+\gamma_-(x;\lambda_1,\lambda_0)\right\} \frac{2\lambda_0}{(1-4x^2)\sqrt{(1-\lambda_0^2)-4x^2}}, 
        \end{equation}
where  
	\begin{equation}\label{gamma}
        \gamma_\pm(x;\lambda_1,\lambda_0)= 
        \frac{1\pm\sqrt{1+ (1-\lambda_1^2)\eta(x)/u^2(x)}+(1-\lambda_1^2)\zeta_\mp(x)/u(x) } {2\left(1+(1-\lambda_1^2)\eta(x)/u^2(x)\right)}.
        \end{equation}
Here  
	\begin{align*} 
        u(x) &= (1-\lambda_0^2)-4x^2, \\ 
        \eta(x) &= 8(1-2\lambda_0^2)x^2-(1+\lambda_1^2-2\lambda_0^2), \\ 
        \zeta_\pm(x) &= 4(2\lambda_0^2-1)x^2+\lambda_1^2\pm \sqrt{\phi(x)}. 
        \end{align*}
We can check from Eq.~(\ref{gamma}) that, for any $x,y\in (-\kappa,\kappa)$, 
\begin{multline*} 
\gamma_+(x;\lambda_1,\lambda_0)=1 \Leftrightarrow \gamma_-(y;\lambda_1,\lambda_0)=0  \\ \Leftrightarrow  \lambda_1=1 \Leftrightarrow \alpha=1. 
\end{multline*}
Thus 
	\[ \rho_*(x;\lambda_1,\lambda_0)=\gamma_-(x;\lambda_1,\lambda_0)\frac{2\lambda_0}{(1-4x^2)\sqrt{(1-\lambda_0^2)-4x^2}} \] 
is the wave which appears  in the weak limit theorem only if the underlying RW is transient. 
We complete the proof. $\square$
\section{Summary}
We considered effects of properties of the underlying RW on the induced Szegedy walk on a magnifier graph. 
The geometric modification to the one-dimensional lattice divides the spectrum into two parts and create the point spectrum. 
First, we showed that the existence of the eigenvalues of the RW provides localization of the induced QW; 
even if the eigenvalue lies in $0$, due to the spectral mapping theorem of the Szegedy walk, the eigenvalue is lifted up to $\pm \im$ 
on the unit circle in $\mathbb{C}$. These eigenspaces are generated by round trip paths. 
Next, we discussed the contribution of continuous spectrum to the induced QW. 
We showed that the pseudo velocity which depends on not only the transience of the underlying RW 
(equivalently, $\lambda_1=\sup|\mathrm{spec}(P)|$), but also the spectral gap (equivalently, $\lambda_0=\inf|\mathrm{spec}(P)\smallsetminus \{0\}|$). 
We can classify $(0,1)^3$, which has a one-to-one correspondence to the underlying RW by Eq.~(\ref{extension}), 
by the relation ``$\sim$" defined as follows: 
for $(p,q,r),\;(p',q',r')\in (0,1)^3$, 
	\[ (p,q,r)\sim (p',q',r') \stackrel{def}{\Longleftrightarrow} \rho(x;p,q,r)=\rho(x;p',q',r'), \]
where $\rho(x;p,q,r)$ is the limit density function of the linear scaled Szegedy walk with the parameters $(p,q,r)$. 
Theorem 2 implies 
that the underlying RW can be non-trivially classified according to the limit behavior of the corresponding QW; that is, 
	\begin{equation*} (p,q,r)\sim (p',q',r') \Longleftrightarrow \lambda_1(p,q,r)=\lambda_1(p',q',r') 
        \;\mathrm{and}\;\lambda_0(p,q,r)=\lambda_0(p',q',r'). 
        \end{equation*}
Here $\lambda_1(p,q,r)$, $\lambda_0(p,q,r)$ are defined by Eqs.~(\ref{pero}) and (\ref{pipi}), 
which are the maximal and minimum of absolute values of the spectrum of the RW, respectively. 
To find the spreading properties of a random walk and its induced quantum walk on $G$, 
in RW case, the maximal absolute value of the underlying spectrum on $[-1,1]$ should be estimated while both of the maximal and minimum ones should be estimated in QW case. 
Thus it suggest that the distribution of the induced QW reflects more detailed properties of the spectrum of the underlying RW 
than the original RW's one. 
The effects of the drifted RW on the induced Szegedy walk in more general setting is one of the interesting future's problem. 

In this paper, we clarified that a homological structure of graph and secondly the point spectrums of the underlying RW are the 
second type of derivations of localization of the Grover walk. 
Recently, we find the third type of the derivation of the localization which reflects a hyperbolicity of the graph. 
This detailed discussion will be seen in the forthcoming paper~\cite{HS}. 

\noindent \\
\noindent \\
{\bf Acknowledgments} 
YuH's work was supported in part by Japan Society for the
Promotion of Science Grant-in-Aid for Scientific Research (C) 25400208 and (B) 24340031 and for Challenging Exploratory Research 26610025. 
ES thanks to the financial supports of the Grant-in-Aid for Young Scientists (B) and of Scientific Research (B) Japan Society for the
Promotion of Science (Grants No.25800088, No.23340027). 

\bibliographystyle{jplain}

\end{document}